# Second-harmonic generation in multilayer hexagonal boron nitride flakes

Sejeong Kim[1,*], Johannes E. Fröch[1], Augustine Gardner[1], Chi Li[1], Igor Aharonovich[1], and Alexander S. Solntsev[1,*]

[1]School of Mathematical and Physical Sciences, University of Technology Sydney, Ultimo, New South Wales, 2007, Australia
*Corresponding authors: Sejeong.Kim-1@uts.edu.au, Alexander.Solntsev@uts.edu.au

**We report second-harmonic generation (SHG) from thick hexagonal boron nitride (hBN) flakes with approximately 109-111 layers. The resulting effective second-order susceptibility is similar to previously reported few-layer experiments. This confirms that thick hBN flakes can serve as a platform for nonlinear optics, which is useful because thick flakes are easy to exfoliate while retaining a large flake size. We also show spatial second-harmonic maps revealing that SHG remains a useful tool for the characterization of the layer structure even in the case of a large number of layers.**

http://dx.doi.org/10.1364/OL.99.099999

Layered materials comprise an extremely promising class of materials for optics and photonics. The most widely studied layered materials are graphene [1], hexagonal boron nitride [2], and transition metal dichalcogenides (TMDCs) such as $MoS_2$, $WS_2$, $MoSe_2$, and $WSe_2$ [3]. TMDCs and hBN have a direct bandgap and show bright photoluminescence [4, 5] and strong second-harmonic generation [6, 7], while SHG in graphene requires special symmetry [8].

Second-harmonic generation is particularly important in this context because it allows quantifying quadratic nonlinear-optical susceptibility ($\chi^{(2)}$) of these layered materials [9-11] for potential integration with mature integrated photonic platforms such as silicon on insulator [12] or silicon nitride on insulator [13]. These platforms have near-zero $\chi^{(2)}$, and thus adding a thin layer of material with strong $\chi^{(2)}$ on top of a photonic structure [14, 15] may enable a range of $\chi^{(2)}$-based applications such as nonlinear spectroscopy [16] and nonlinear signal processing [17, 18].

Furthermore, SHG can be used to extract specific information from layered materials which is non-trivial for other existing techniques. For example, SHG intensity is strong for odd numbers of layers while it goes to zero for even numbers of layers due to interference [19]. Therefore, the intensity of the SHG signal can be used to determine an even or an odd number of layers. Additionally, SHG intensity depends on the orientation of material vs. the excitation polarization [20, 21]. Thus nonlinear spectroscopy of layered materials is a powerful tool to determine the crystalline orientation, thickness uniformity, and layer stacking [6]. One of the biggest advantages of using hBN is its transparency in the entire visible range combined with a relatively high refractive index ($n$ = 2.1 at 600 nm) for transparent materials [22], potentially opening doors for visible nonlinear optical applications [23]. The SHG in hBN has been studied [19, 20, 24, 25], however less intensively investigated compared to TMDCs, likely due to lower $\chi^{(2)}$ [7].

In this work, we experimentally confirm that, despite predictions from earlier 1-to-5 layers experiments [19] indicating a potential gradual decrease of SHG efficiency with increasing hBN thickness, strong SHG can be achieved in hBN with above 100 layers, enabling the use of thick mechanically exfoliated hBN lakes in nonlinear optics.

The optical measurement setup used for the characterization is shown in Fig. 1. A femtosecond pulsed laser with 80 MHz repetition rate is used as an excitation laser. The width of the pulse is measured to be 80 fs. The pump laser is focused through a 50 x objective lens with a numerical aperture of 0.55, and the same objective lens is used to collect the second-harmonic signal from the hBN flake. The hBN sample is mounted on the piezo-stage, and the second-harmonic signal is measured by a spectrometer (QEpro, OceanOptics) to show spectra and 2D mapping.

The system is calibrated using the laser with the same wavelength as the second-harmonic signal, to convert intensity values measured at the spectrometer to the power generated by the sample. 405 nm laser shown in Fig. 1 is attenuated by the ND filter and directly measured by the spectrometer. The transmission of the objective lens at 405 nm is also measured for the system calibration.

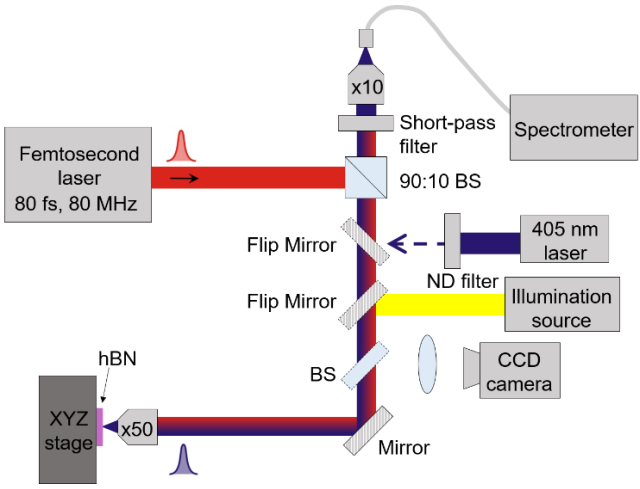

**Fig. 1.** Custom-built experimental setup for optical characterization. A femtosecond laser is used to measure the second-harmonic response of hexagonal Boron Nitride (BS: beam splitter, ND filter: neutral-density filter).

For sample preparation, the hBN bulk crystal is mechanically exfoliated on a SiO$_2$ substrate. The sample is heated in air at 500°C and subsequently annealed in vacuum at 850°C for 30 minutes to remove the polymer residuals from sticky tapes. An optical microscope image of the transferred flake is shown in Fig. 2(a). The same hBN flake is analyzed using atomic force microscopy (AFM) and Raman spectroscopy. Fig. 2(b) shows that the thickness of the flake is 36.2 nm. It was demonstrated previously that an interlayer distance of 0.33 nm could be used to estimate the number of layers in hBN [26]; therefore, the flake shown in Fig. 2 contains approximately 110±2 layers. Based on the previously discussed observation that an odd number of layers is expected to allow strong SHG, the flake is likely to contain 109 or 111 layers. Raman spectroscopy is used to identify hBN crystalline structure showing the characteristic hBN E$_{2g}$ Raman mode at 1366 cm$^{-1}$ as shown in the inset in Fig. 2(a). Figure 2(c) shows the pump spectrum with the central wavelength of 810 nm and a linewidth of 10.2 nm. The second-harmonic signal is observed at 405 nm, with a reduced linewidth of 4.1 nm, as shown in Fig. 2(d). The line narrowing between the pump and the SHG signal of approximately $2\sqrt{2}$ times is consistent with predictions based on conventional frequency doubling theory in the absence of dispersion [27]. The gray line in Fig. 2(d) shows the negligible nonlinear response from the SiO$_2$ substrate.

Figure 3(a) shows power-dependent spectra of SHG from the hBN sample characterized in Fig. 2. The signal is collected from the center of the flake to avoid the edge effect. The excitation power in Fig. 3(a) is measured before the objective lens. To plot the pump-power-dependent SHG signal to demonstrate the quadratic behavior, the values in Fig. 3(a) are carefully converted. First, the excitation power onto the hBN sample is derived from the power measured before objective multiplied by the transmission of the objective lens at 405 nm, which in our case is 0.709. Next, the integrated intensity from spectra plotted with arbitrary units is converted into real power using the calibration with a 405 nm laser. The measurements are performed with 3-second integration time. The Log-log plot [Fig. 3(b)] of the excitation power and SHG power showing a linear slope of 2.05 ± 0.126.

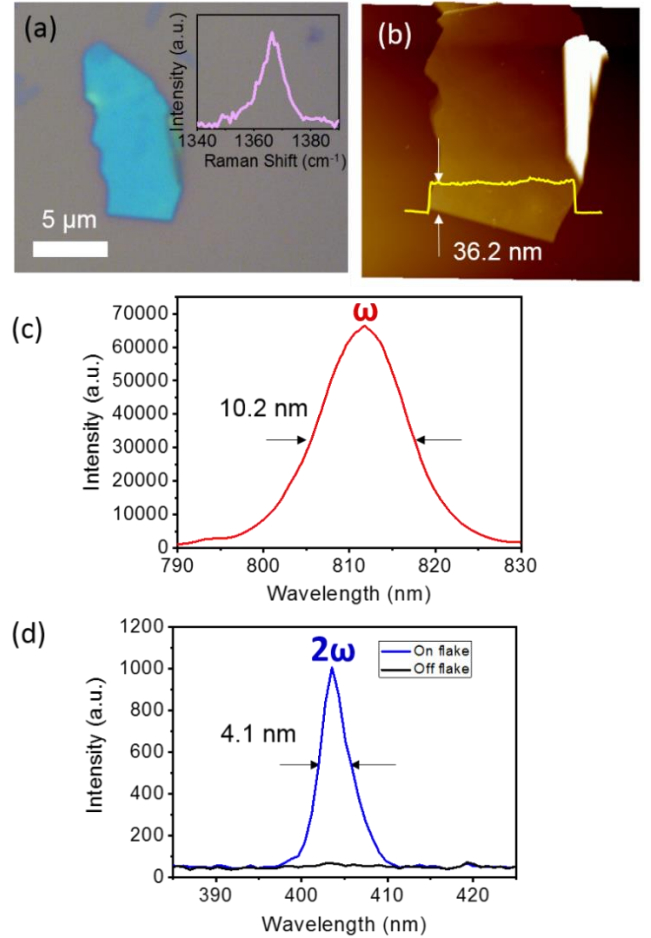

**Fig. 2.** Second-harmonic generation from an hBN flake. (a) Optical image of the studied mechanically exfoliated hBN flake with Raman spectrum of the hBN flake in the inset. (b) Atomic force microscopy (AFM) map of the flake showing the thickness of flake corresponding to 36.2 nm. (c) The pump laser wavelength at 810 nm with full-width half-maximum (FWHM) of 10.2 nm. (d) SHG from hBN flake (solid blue line) with FWHM of 4.1 nm. The solid black line shows the negligible signal from the SiO$_2$ substrate.

The second-order susceptibility of the studied hBN flake adjusted for its thickness is calculated using the following equation [10]:

$$\left|\chi_{th}^{(2)}\right| = \sqrt{\frac{P(2\omega) \times c \times \varepsilon_0 \times RR \times A \times \Delta\tau \times \lambda^2 \times (1+n)^6}{64\sqrt{2}\pi^2 \times P^2(\omega) \times S \times n^3}}$$

Here $P(2\omega)$ is the average power from the second-harmonic signal, $P(\omega)$ is the average power of the excitation laser, c is the speed of light in vacuum and $\varepsilon_0$ is the vacuum electric permittivity ($8.85 \times 10^{-12}$ F/m). $RR$ is the repetition rate of the excitation laser (80 MHz), $A$ is the pump spot area ($3.14 \times 10^{-12}$ m$^2$), $\Delta\tau$ is the pulse full width at half maximum, and $S$ = 0.94 is a shape factor for Gaussian pulses. We use the refractive index of n = 2.1 [22].

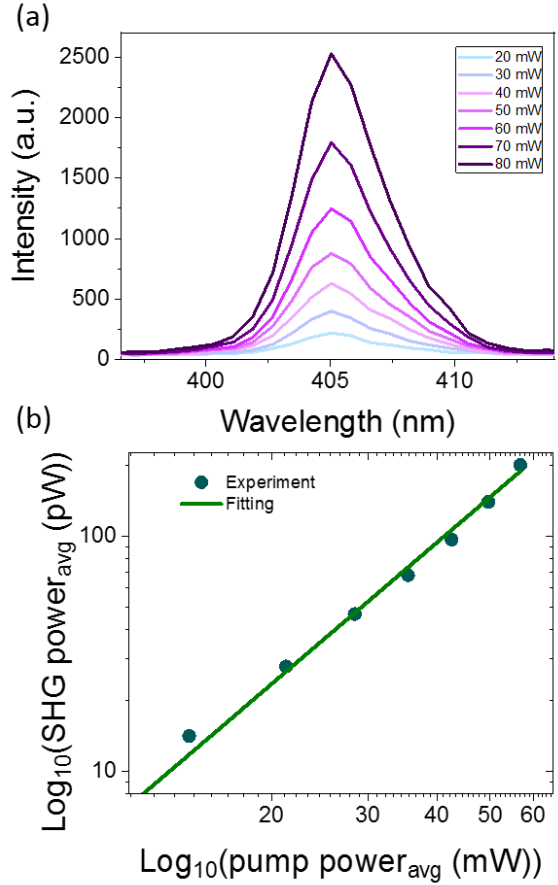

**Fig. 3.** Power-dependent SHG response from hBN. (a) Excitation power-dependent SHG spectra. The pump power is measured before the objective lens. Each spectrum is integrated from 400 nm to 410 nm. The integrated power measured in arbitrary units is converted into exact SHG power. (b) SHG power vs. the excitation pump power on the sample.

The pump spot area ($A$) is identified from the CCD camera. From the equation shown above, we obtain a value of $1.4 \times 10^{-20}$ m$^2$/V. If we assume that only one layer with a thickness $d = 0.33$ nm is responsible for the SHG while other layers interfere destructively as suggested in earlier works [6, 19], then we obtain a value of a single-layer quadratic nonlinear optical susceptibility $\chi^{(2)} = \chi^{(2)}_{th}/d = 4.2 \times 10^{-11}$ m/V, which is similar to the previously reported value for a single layer despite the expectations that quadratic nonlinearity should decrease when the number of layers increase [19]. A likely explanation for this phenomenon is a difference in the quality of the utilised hBN flakes.

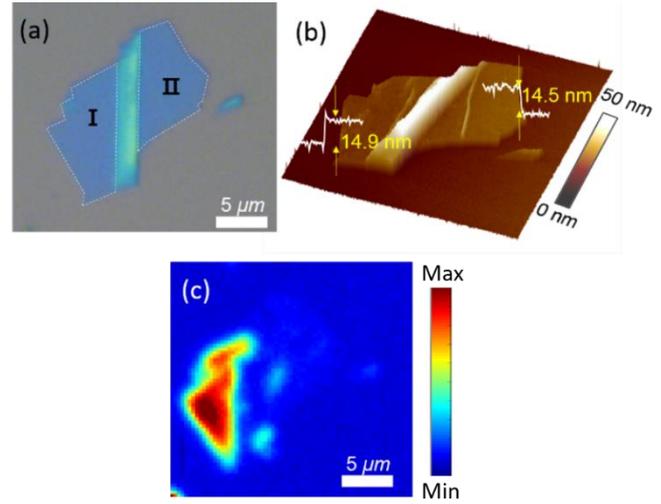

**Fig. 4.** Spatial mapping. (a) Optical image of hBN flake with areas of different layer number labeled as I and II. (b) AFM map of the flake shown in (a). (c) Spatial SHG intensity mapping of the hBN sample.

Due to the layered nature of hBN, a flake may have non-homogeneous thickness. Therefore, it is important to map the SHG from a flake. An hBN flake shown in Fig. 4(a) is used to acquire a second-harmonic map. A larger and thinner flake is chosen for mapping because it is easier to obtain a more precise AFM map of a flake with the lower thickness in our apparatus and simultaneously utilize a larger scanning area to build a more detailed SHG map. The flake is analyzed using AFM and exhibits areas of different thickness, on the left (region I) and the right (region II) side separated by the folded line in the center, as shown in Fig. 4(b).

To construct an SHG 2D map [Fig. 4(c)], the sample is scanned with a piezo-stage while collecting a spectrum at each step. Subsequently, these spectra are post-processed to extract the intensity at the SHG wavelength. The excitation wavelength of 890 nm is used to pump the sample. The wavelength has been chosen to optimize the transmission throughout the setup, including objectives and filters, to allow faster scans. Each pixel in the 2D map shows the intensity at 445 nm. The excitation power measured before the objective lens is 40 mW. The collection time for each spectrum is 500 ms. This sample-scanning spectrometer-based approach can avoid the possible noise that may affect conventional SHG microscopy that uses a CCD camera and a bandpass filter to construct the SHG 2D image. Region I shows strong SHG signal, while the signal from region II is negligible. This is because hBN layers are stacked in 2H order, where the sites for boron and nitrogen alternating when the layer increases. Therefore, the inversion symmetry is only broken for an odd number of layers. Based on the SHG mapping and the AFM analysis, region I contains an odd number of stacked layers (likely 45 layers with a thickness of 14.9 nm), which is 1 layer thicker than region II (likely 44 layers with a thickness of 14.5 nm). Note that similar to TMDSs and Graphene, hBN flakes mechanically exfoliated from a high-quality crystal can have a negligible surface roughness (with minor nanoscopic surface defects) due to strong covalent bonds between atoms within the layer and week Van der Waals bonds between the layers. This, however, does not necessarily hold true for the edges of the flake, potentially leading to the non-uniformity in the SHG map.

In summary, we have demonstrated strong SHG from thick hBN flakes with above 100 layers. We have measured the normalized quadratic nonlinear optical susceptibility and confirmed that it is comparable to previously reported measurements in single hBN layers. We have also demonstrated with spatial mapping that SHG may be a useful tool to complement the characterization of the layer structure of thick flakes.

In the future, it will be interesting to test systems with intermediate thickness where the number of layers can be measured more precisely, perform experiments with large deterministically controlled numbers of layers, and to integrate hBN flakes with microstructures [14, 15, 28]. Notably, integration with micro-structures will require careful design that takes into account the phase-matching considerations. In this paper, as well as in most previous works on nonlinear optics in 2D materials [6, 7], phase-matching does not play a role because of the strongly sub-wavelength interaction length, which would not hold true for waveguide integration [15]. Other potentially interesting research directions include experimental verification of theory for hBN SHG in deep-UV [24] and SHG in hBN nano-structures [25].

**Funding.** Australian Research Council (ARC) (DE180100070, DP190101058).

**Disclosures.** The authors declare no conflicts of interest.